\documentclass[final]{aipproc}

\layoutstyle{6x9}
\usepackage{units}
\usepackage{xspace}
\usepackage[latin1]{inputenc}

\begin{document}

\newcommand{\as}{\ensuremath{\alpha_s}\xspace}
\newcommand{\asmz}{\ensuremath{\alpha_s(M_Z)}\xspace}
\newcommand{\xbj}{\ensuremath{x_{\rm{Bj}}}\xspace}
\newcommand{\Qsq}{\ensuremath{Q^2}\xspace}

\newcommand{\muf}{\ensuremath{\mu_{\mathrm{f}}}\xspace}
\newcommand{\mur}{\ensuremath{\mu_{\mathrm{r}}}\xspace}

\newcommand{\MeanPt}{\ensuremath{\langle P_{\mathrm{T}} \rangle}\xspace}
\newcommand{\Pt}{\ensuremath{P_{\mathrm{T}}}\xspace}
\newcommand{\etalab}{\ensuremath{\eta_{\mathrm{lab}}}\xspace}
\newcommand{\Mjj}{\ensuremath{M_{\mathrm{jj}}}\xspace}
\newcommand{\xij}{\ensuremath{\xi}\xspace}

\newcommand{\Pth}{\ensuremath{P_{\mathrm{T}}^{\mathrm{h}}}\xspace}
\newcommand{\Ptda}{\ensuremath{P_{\mathrm{T}}^{\mathrm{da}}}\xspace}

\newcommand{\NLOJet}{\textsc{nlojet}\texttt{++}\xspace}
\newcommand{\FastNLO}{\textsc{fastnlo}\xspace}
\newcommand{\Ariadne}{\textsc{ariadne}\xspace}
\newcommand{\Jetset}{\textsc{jetset}\xspace}
\newcommand{\Pythia}{\textsc{pythia}\xspace}
\newcommand{\Rapgap}{\textsc{rapgap}\xspace}
\newcommand{\Lepto}{\textsc{lepto}\xspace}
\newcommand{\Djangoh}{\textsc{djangoh}\xspace}
\newcommand{\QCDNUM}{\textsc{qcdnum}\xspace}	

\newcommand{\HERAI}{HERA-1\xspace}
\newcommand{\HERAII}{HERA-2\xspace}

\title{Measurement of Multijet Production in DIS and Determination of the Strong Coupling Constant}

\classification{13.87.-a, 13.87.Ce, 12.38.Qk}
\keywords      {Jet production, Quantum Chromodynamics, Strong coupling constant}

\author{R. Kogler}{
  address={\vskip-1mm(for the H1 Collaboration) \\
Deutsches Elektronen Synchrotron, Notkestra\ss{}e 85, 22607 Hamburg, Germany}
}

\begin{abstract}
Inclusive jet, dijet and trijet differential cross sections have been measured in neutral current deep-inelastic scattering for exchanged boson virtualities $150 < \Qsq < \unit[15000]{GeV^2}$ with the H1 detector at HERA using an integrated luminosity of $\unit[351]{pb^{-1}}$. The multijet cross sections are presented as a function of \Qsq, the transverse momentum of the jet \Pt (the mean transverse momentum for dijets and trijets) and the proton's longitudinal momentum fraction of the parton participating in the hard interaction \xij. The cross sections are compared to perturbative QCD calculations at next-to-leading order and the value of the strong coupling \asmz is determined. 
\end{abstract}

\maketitle


\section{Introduction}

Jet production in neutral current (NC) deep-inelastic scattering (DIS) provides an ideal environment for studying Quantum Chromodynamics (QCD). While inclusive DIS gives only indirect information on the strong coupling, \as, via scaling violations of the proton structure functions, the production of jets allows a direct measurement of \as. 

The Born level contribution to DIS generates no transverse momentum in the Breit frame, where the virtual boson and the proton collide head on. Significant transverse momentum \Pt in the Breit frame is produced at leading order (LO) in \as by the QCD-Compton and boson-gluon fusion processes. In LO the proton's momentum fraction carried by the emerging parton is given by $\xij =  \xbj ( 1+ \Mjj^2 / \Qsq)$, where \xbj denotes the Bjorken scaling variable, \Mjj the invariant mass of the two jets of highest \Pt and \Qsq the negative four-momentum transfer squared. In the kinematical region of low \Qsq and low \Pt boson-gluon fusion dominates jet production and provides direct sensitivity to the gluon component of parton density functions (PDFs). At high \Qsq and high \Pt QCD-Compton processes are dominant, which are sensitive to the valence quark distributions. The inclusion of jet cross sections into the extraction of PDFs thus provides an important constraint on \as and disentangles the correlation between \as and the gluon \cite{Nowak:2011}. 


\section{Multijet Measurement}

In a recent publication on multi-jet production at high \Qsq by the H1 collaboration~\cite{Aaron:10:363} measurements of normalised multi-jet cross sections and extractions of \asmz have been made exploiting the full \HERAI and \HERAII data sets. Measurements of absolute cross sections of inclusive jet production at high \Qsq have been published by the H1 collaboration~\cite{Aktas:07:134} using \HERAI data only. In the analyses presented here jet cross sections are measured making use of the full \HERAII data with an integrated luminosity of \unit[351]{pb}$^{-1}$. Improvements in the reconstruction of tracks and calorimetric energy have been made, which lead to a better jet energy resolution and a jet energy scale uncertainty of 1\%, which improves the precision of the jet cross section measurement by up to 50\% with respect to previous publications.

\begin{figure}
\centering
 \includegraphics[height=7cm]{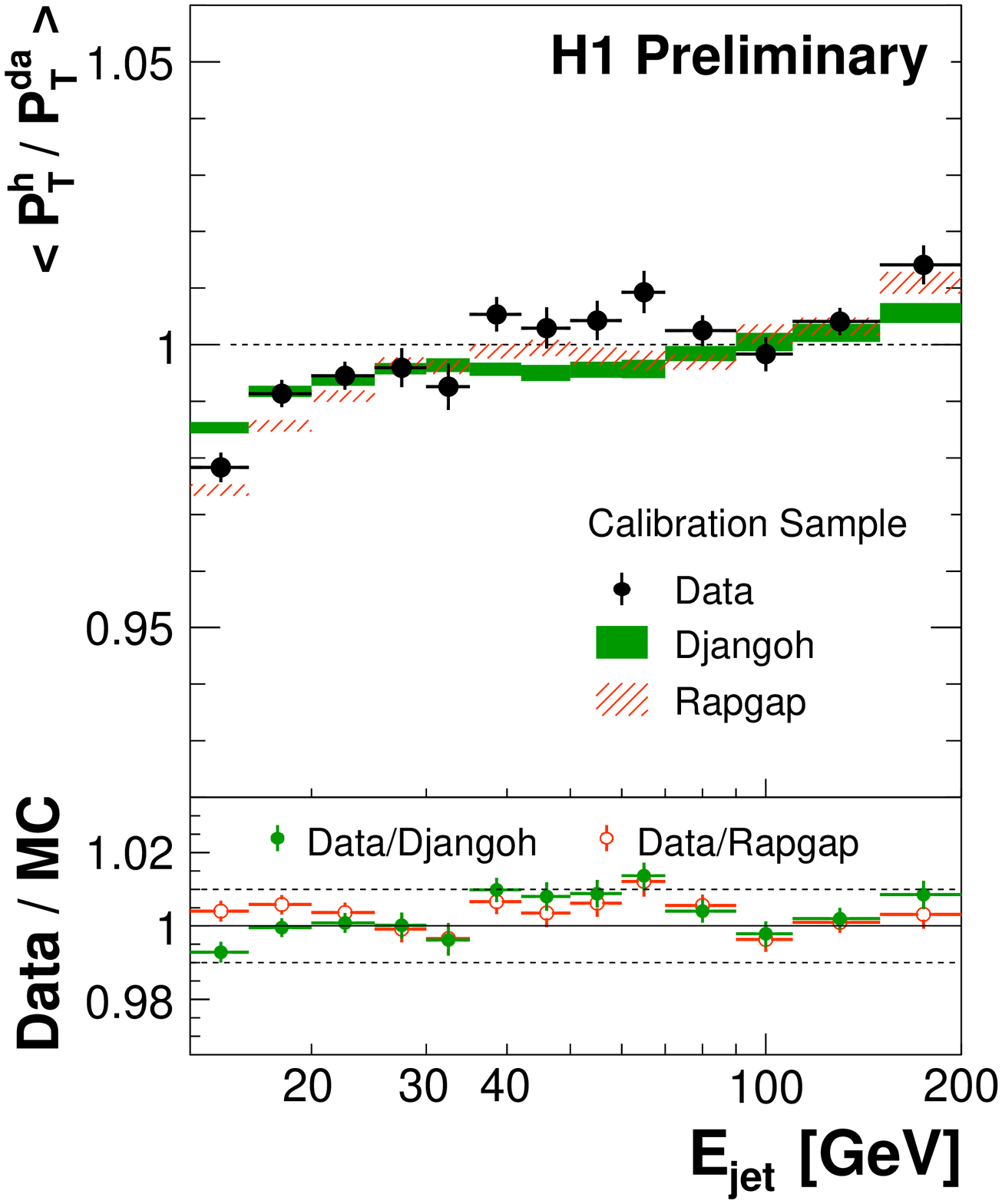} \put(-123,177){\footnotesize (a)}\hspace{0.3cm}
 \includegraphics[height=7cm]{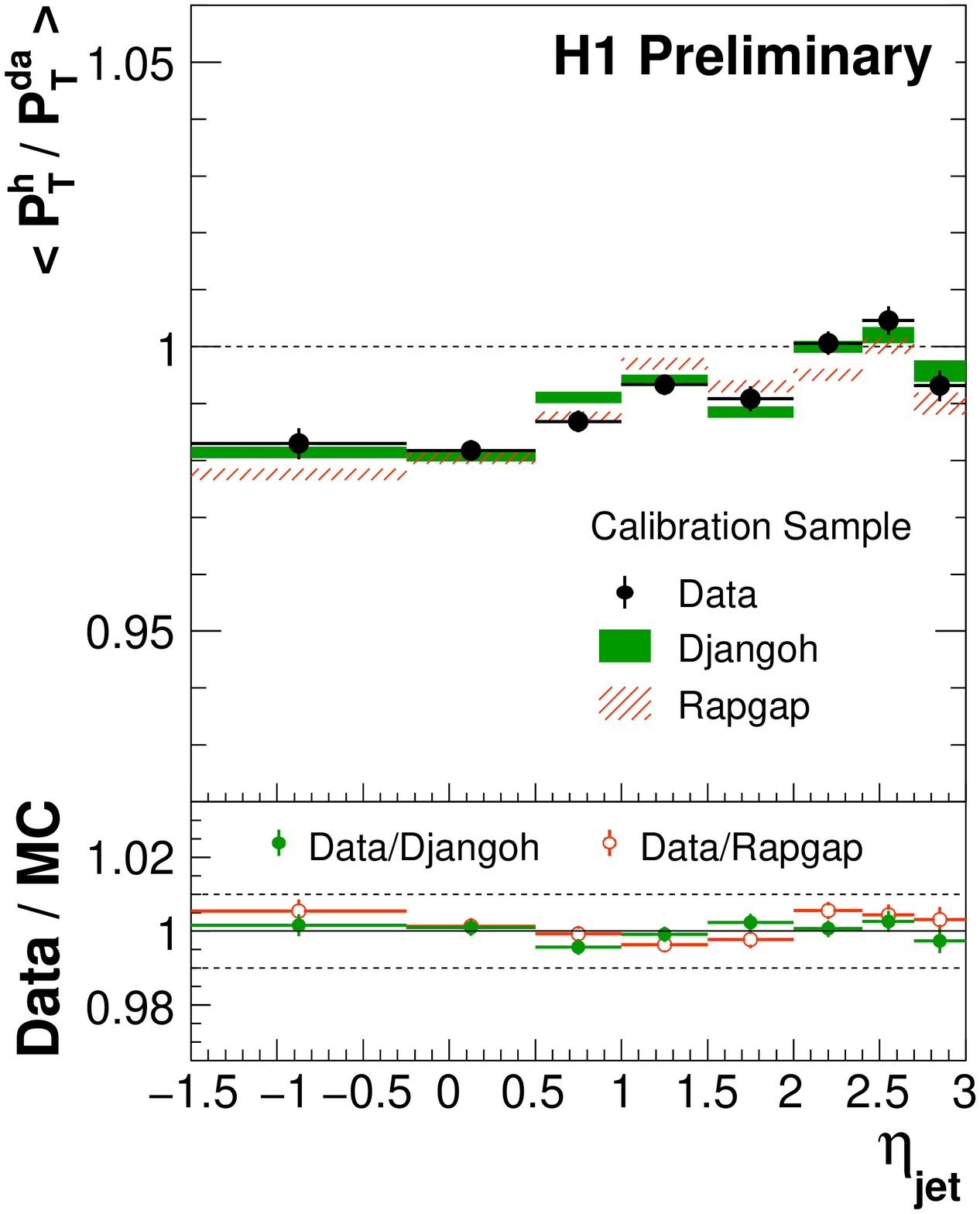} \put(-123,177){\footnotesize (b)}
\caption{Ratio of the reconstructed and calibrated transverse momentum of the HFS \Pth, to the reference measurement \Ptda as function of the jet energy (a) and jet pseudorapidity (b). The double-ratio of data to MC is shown at the bottom of each plot. 
\label{fig:calib}
}
\end{figure}
The NC DIS events, which form the basis of the jet analyses, are primarily selected by requiring a scattered electron in the main liquid Argon (LAr) calorimeter of H1. They have to fulfil $150 < \Qsq < \unit[15000]{GeV^2}$ and $0.2 < y < 0.7$, where $y$ refers to the inelasticity of the interaction. The jet finding is performed in the Breit frame, where the boost from the laboratory system is determined by \Qsq, $y$ and by the azimuthal angle of the scattered electron, $\phi_e$. Particles of the hadronic final state (HFS) are clustered into jets using the inclusive $k_\mathrm{T}$ algorithm \cite{Ellis:93:3160} as implemented in FastJet \cite{Cacciari:06:57} with the massless \Pt recombination scheme and with the distance parameter $R_0 = 1$. The requirement \mbox{$-1.0 < \etalab < 2.5$}, where \etalab is the jet pseudorapidity in the laboratory frame, ensures that jets are contained within the acceptance of the LAr calorimeter and are well calibrated. Jets are ordered by decreasing transverse momentum \Pt in the Breit frame, which is identical to the transverse energy for massless jets. Every jet with the transverse momentum \Pt in the Breit frame satisfying $7 < \Pt < \unit[50]{GeV}$ contributes to the inclusive jet cross section. Events with at least two (three) jets with transverse momentum $5 < \Pt < \unit[50]{GeV}$ are considered as dijet (trijet) events. In order to avoid regions of phase-space where fixed order perturbation theory is not reliable, dijet events are accepted only if the invariant mass \Mjj of the two leading jets exceeds \unit[16]{GeV}. The same requirement on \Mjj is applied to the trijet events so that the trijet sample is a subset of the dijet sample.


For the reconstruction of the HFS an energy-flow algorithm is used by H1, which combines information from tracking and calorimetric measurements. A detailed description of the employed track detectors and the LAr calorimeter can be found elsewhere \cite{Abt:97:310}. Recently a large effort has been made by the H1 collaboration to improve on all aspects of the reconstruction of the HFS. The track and vertex reconstruction has been improved by using a double-helix trajectory, which takes into account secondary scatterings inside the detector material. The calorimetric measurement benefits from a separation of hadronic and electromagnetic showers based on shower shape estimators and neural networks \cite{Kogler:2011}. Based on these improvements a new calibration method for the HFS has been developed, allowing the calibration of individual calorimetric deposits depending on their composition of hadronic and electromagnetic components \cite{Kogler:2011}.

Figure \ref{fig:calib} shows the mean value of the ratio of \Pth, which is the calibrated transverse momentum of the HFS, to \Ptda, which is the transverse momentum of the HFS using the double-angle method \cite{Bassler:95:197}. The variable \Ptda is independent of the calibration and is used as a reference scale in both, data and MC simulations. The reconstruction of the absolute energy scale is achieved within 2\%, where fluctuations can be due to the method of obtaining the mean value of the distributions of $\Pth / \Ptda$. At the bottom of Fig.~\ref{fig:calib} the double-ratio of data to MC is shown. A jet energy scale uncertainty of only 1\% is achieved over the full accessible range of jet energy $E_{\mathrm{jet}}$ and pseudorapidity $\eta_{\mathrm{jet}}$.

\begin{figure}
\centering
 \includegraphics[height=4.5cm]{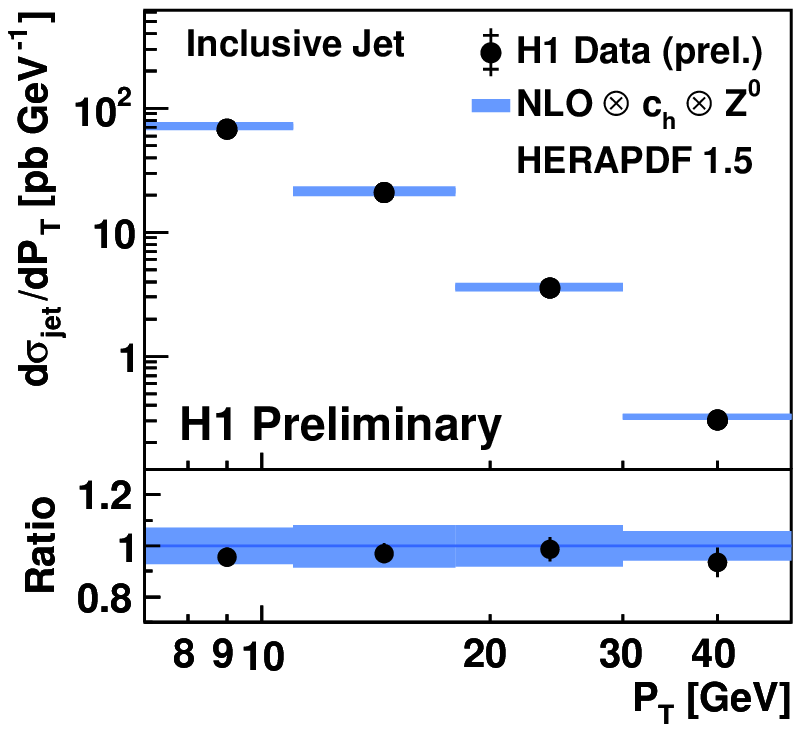} \put(-108,62){\footnotesize (a)} \hspace{0.08cm}
 \includegraphics[height=4.5cm]{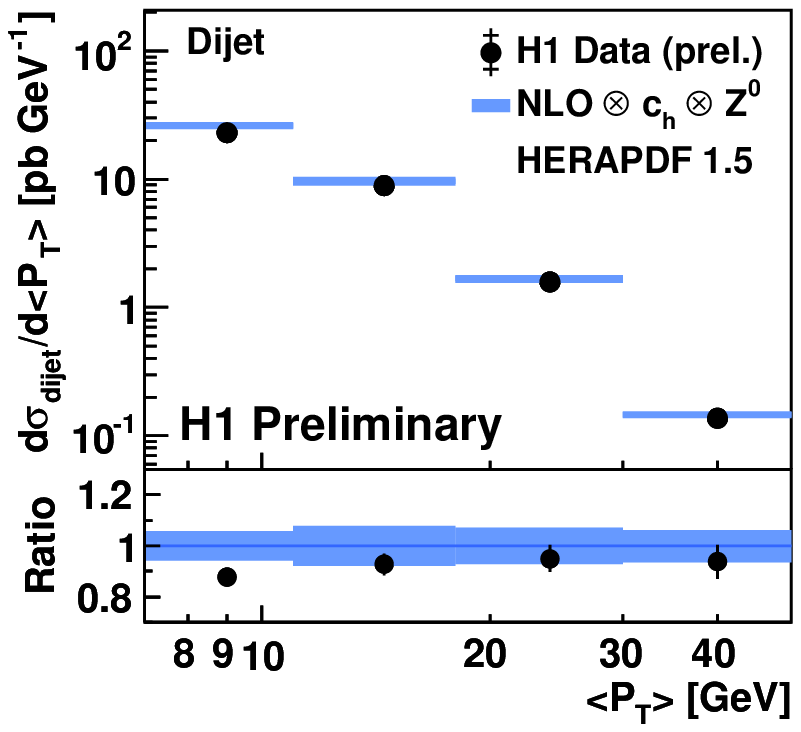} \put(-108,62){\footnotesize (b)} \hspace{0.08cm}
 \includegraphics[height=4.5cm]{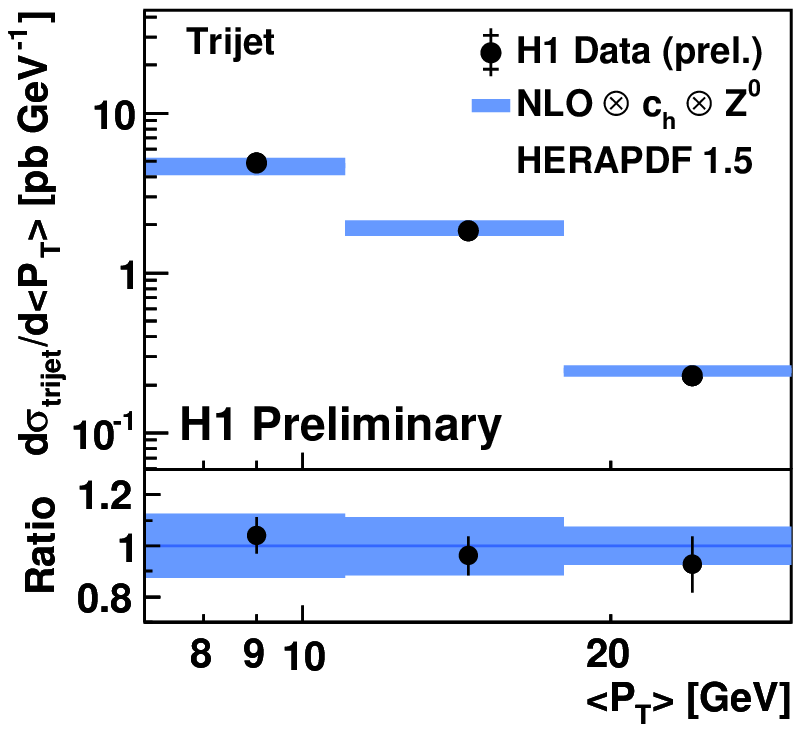} \put(-108,62){\footnotesize (c)}
\caption{Inclusive jet (a), dijet (b) and trijet (c) cross sections measured as function of \Pt of the jet (\MeanPt in case of the dijet and trijet measurements). The measurement is compared to NLO pQCD calculations, which have been corrected for hadronisation effects and effects from $Z^0$ exchange. 
\label{fig:xs}
}
\end{figure}

The measured jet cross sections as function of \Pt and \MeanPt are presented in Fig.~\ref{fig:xs}. They are compared to NLO perturbative QCD calculations obtained with \NLOJet~\cite{Nagy:99}. The calculations are obtained using the HERAPDF1.5~\cite{HERAPDF1.5} parametrisations, with the factorisation and renormalisation scales set to $\muf = \mur = \sqrt{(\Pt^2+\Qsq)/2}$, where \MeanPt is used instead of \Pt in the case of the dijet and trijet measurements. The NLO calculations are corrected for hadronisation effects and effects from $Z^0$ exchange, which are not included in \NLOJet. The uncertainties of the theoretical predictions due to missing higher orders are estimated by varying \muf and \mur independently up and down by a factor of two. 
The uncertainties on the measured cross sections are about a factor of two smaller than the theoretical uncertainties, which demonstrates the precision of the presented jet measurement. The inclusive jet, dijet and trijet cross sections are also measured double-differentially as function of \Qsq and \Pt and as function of \Qsq and \xij \cite{H1Prel-11-032}. These jet data will provide important input for future PDF determinations, where they can help to disentangle the correlation between the gluon and \as. In order to fully exploit these results next-to-next-to-leading order calculations or resummations of logarithms are required. 

\section{Determination of the Strong Coupling Constant}

The inclusive jet, dijet and trijet cross sections are used to extract the value of \asmz. The extraction is performed with fits based on a $\chi^2$-minimisation, which takes correlated and uncorrelated uncertainties fully into account \cite{Aaron:10:363, Aktas:07:134}. During the fitting procedure the theoretical predictions are obtained with \FastNLO \cite{Kluge:06}, which uses \NLOJet. The values extracted from the inclusive jet, dijet and trijet measurements are: \vspace{-0.2cm}
\begin{eqnarray*}
\mathrm{inc.~jet:}~\alpha_s(M_Z) = 0.1190 \pm 0.0021 ~\mathrm{(exp.)} \pm 0.0020 ~\mathrm{(pdf)}^{~+0.0050}_{~-0.0056}~\mathrm{(th.)} \\
\mathrm{dijet:}~\alpha_s(M_Z) = 0.1146 \pm 0.0022 ~\mathrm{(exp.)} \pm 0.0021 ~\mathrm{(pdf)}^{~+0.0044}_{~-0.0045}~\mathrm{(th.)} \\
\mathrm{trijet:}~\alpha_s(M_Z) = 0.1196 \pm 0.0016 ~\mathrm{(exp.)} \pm 0.0010 ~\mathrm{(pdf)}^{~+0.0055}_{~-0.0039}~\mathrm{(th.)} \\[-0.8cm]
\end{eqnarray*}
which are compatible within the uncertainties. These values are consistent with previously determined values of \asmz by H1 \cite{Aaron:10:363, Aktas:07:134} and the world average \cite{Bethke:09:689}. The experimental uncertainties are somewhat larger than in the previous H1 determination from high \Qsq jet data, where normalised jet cross sections have been used. This leads to smaller experimental uncertainties since all normalisation uncertainties cancel in the ratio. In the case of the \as extraction from inclusive jet and dijet data the normalisation uncertainties give the largest contribution to the experimental uncertainty. In the trijet case the experimental uncertainty is smallest, since the cross section is $\mathcal{O}(\as^2)$ already in LO and therefore more sensitive to the slope of the cross section and less affected by the overall normalisation of the data. Also in the trijet case, the value of \asmz has the smallest PDF uncertainty. In all cases the theoretical uncertainties, which are dominated by uncertainties due to missing higher orders, are larger by about a factor of two than the experimental and PDF uncertainties.



\bibliographystyle{aipproc}   

\bibliography{ShortReferences}

\begin{thebibliography}{13}
\expandafter\ifx\csname natexlab\endcsname\relax\def\natexlab#1{#1}\fi
\providecommand{\enquote}[1]{``#1''}
\expandafter\ifx\csname url\endcsname\relax
  \def\url#1{\texttt{#1}}\fi
\expandafter\ifx\csname urlprefix\endcsname\relax\def\urlprefix{URL }\fi
\providecommand{\eprint}[2][]{\url{#2}}

\bibitem[Nowak(2011)]{Nowak:2011}
K.~Nowak, \emph{{these proceedings, also H1prelim-11-034, ZEUS-prel-11-001}}
  (2011).

\bibitem[{F.~D.~Aaron {\it et al.} (H1 Collaboration)}(2010)]{Aaron:10:363}
{F.~D.~Aaron {\it et al.} (H1 Collaboration)}, \emph{Eur. Phys. J. C}
  \textbf{65}, 363 (2010).

\bibitem[{A Aktas {\it et al.} (H1 Collaboration)}(2007)]{Aktas:07:134}
{A Aktas {\it et al.} (H1 Collaboration)}, \emph{Phys. Lett. B} \textbf{653},
  134 (2007).

\bibitem[Ellis and Soper(1993)]{Ellis:93:3160}
S.~D. Ellis, and D.~E. Soper, \emph{Phys. Rev. D} \textbf{48}, 3160 (1993).

\bibitem[Cacciari and Salam(2006)]{Cacciari:06:57}
M.~Cacciari, and G.~P. Salam, \emph{Phys. Lett. B} \textbf{641}, 57 (2006).

\bibitem[{I Abt {\it et al.} (H1 Collaboration)}(1997)]{Abt:97:310}
{I Abt {\it et al.} (H1 Collaboration)}, \emph{Nucl. Instr. and Meth. A}
  \textbf{386}, 310 (1997).

\bibitem[Kogler(2011)]{Kogler:2011}
R.~Kogler, Ph.D. thesis, Universität Hamburg, DESY-THESIS-2011-003,
  MPP-2010-175 (2011).

\bibitem[Bassler and Bernardi(1995)]{Bassler:95:197}
U.~Bassler, and G.~Bernardi, \emph{Nucl. Instr. and Meth. A} \textbf{361}, 197
  (1995).

\bibitem[Nagy and Trocsanyi(1999)]{Nagy:99}
Z.~Nagy, and Z.~Trocsanyi, \emph{Phys. Rev. D} \textbf{59}, 14020 (1999), {\it
  Phys.\ Rev.\ Lett.\ }{\bf 87}, 82001 (2001).

\bibitem[{H1 and ZEUS Collaborations}(2010)]{HERAPDF1.5}
{H1 and ZEUS Collaborations}, \emph{preliminary result: H1prelim-10-142,
  ZEUS-prel-10-018}  (2010).

\bibitem[{H1 Collaboration}(2011)]{H1Prel-11-032}
{H1 Collaboration}, \emph{preliminary result: H1prelim-11-032}  (2011).

\bibitem[Kluge et~al.(2006)]{Kluge:06}
T.~Kluge, K.~Rabbertz, and M.~Wobisch  (2006), \eprint{hep-ph/0609285}.

\bibitem[Bethke(2009)]{Bethke:09:689}
S.~Bethke, \emph{Eur. Phys. J. C} \textbf{64}, 689 (2009).

\end{thebibliography}

\end{document}